\documentclass[twocolumn,showpacs,preprintnumbers,amsmath,amssymb]{revtex4}

\usepackage{graphicx}% Include figure files
\usepackage{dcolumn}% Align table columns on decimal point
\usepackage{bm}% bold math

\begin{document}
\newcommand{\braopket}[3]{\langle #1 | \hat #2 |#3\rangle}
\newcommand{\braket}[2]{\langle #1|#2\rangle}
\newcommand{\bra}[1]{\langle #1|}
\newcommand{\braketbraket}[4]{\langle #1|#2\rangle\langle #3|#4\rangle}
\newcommand{\braop}[2]{\langle #1| \hat #2}
\newcommand{\ket}[1]{|#1 \rangle}
\newcommand{\ketbra}[2]{|#1\rangle \langle #2|}
\newcommand{\op}[1]{\hat {#1}}
\newcommand{\opket}[2]{\hat #1 | #2 \rangle}
\preprint{APS/123-QED}

\title{Indirect exchange coupling between localized magnetic moments in carbon nanotubes: a dynamic approach}% Force line breaks with \\

\author{A. T. Costa}
\email{antc@if.uff.br}
\affiliation{
Instituto de F\'isica, Universidade Federal Fluminense, 24210-346 Niter\'oi, RJ, Brazil
}
\author{R. B. Muniz}
\affiliation{
Instituto de F\'isica, Universidade Federal Fluminense, 24210-346 Niter\'oi, RJ, Brazil
}

\author{M. S. Ferreira}
\email{ferreirm@tcd.ie}
\affiliation{
School of Physics, Trinity College Dublin, Dublin 2, Ireland
}

\date{\today}% It is always \today, today,
% but any date may be explicitly specified

\begin{abstract}

Magnetic moments dilutely dispersed in a metallic host tend to be coupled through the conduction electrons of the metal. This indirect exchange coupling, known to occur for a variety of magnetic materials embedded in several different metallic structures, is of rather long range, especially for low-dimensional structures like carbon nanotubes. Motivated by recent claims that the indirect coupling between magnetic moments in precessional motion has a much longer range than its static counterpart, here we consider how magnetic atoms adsorbed to the walls of a metallic nanotube respond to a time-dependent perturbation that induces their magnetic moments to precess. By calculating the frequency-dependent spin susceptibility we are able to identify resonant peaks whose respective widths provide information about the dynamic aspect of the indirect exchange coupling. We show that by departing from a purely static representation to another in which the moments are allowed to precess, we change from what is already considered a long range interaction to another whose range is far superior. In other words, localized magnetic moments embedded in a metallic structure can feel each other's presence more easily when they are set in precessional motion. We argue that such an effect can have useful applications leading to large-scale spintronics devices. 

\end{abstract}

\maketitle
\bibliographystyle{apsrev} % Choose Phys. Rev. style for bibliography

\section{\label{sec:level1}\protect Introduction}

The demand for faster computers has transformed the field of
magnetoelectronics from a purely academic field of research into one of the most promising scientific areas regarding potential technological applications. When combined with systems of reduced dimensionality like thin films, nanowires and molecular structures, this field of research is expected to pave the way to the production of non-volatile computer memories, efficient magnetic sensors and magnetic materials with enhanced information storage capacity, to name but a few. 

Among low-dimensional structures, carbon nanotubes are potentially useful for magneto-transport applications. The ability to produce sizeable changes in the conductance of a nanotube due to an applied magnetic field is one of the driving forces in the research of magnetic
properties of carbon-based structures\cite{alphenaar, littlewood}.  In this case, nanotubes must interact with magnetic foreign objects that lift the intrinsic spin balance of a non-magnetic material like carbon. Substrates\cite{ferreira04, cespedes04}, substitutional impurities\cite{fazzio1}, adsorbed atoms\cite{fazzio2, fazzio3} and nanoparticles\cite{yang03} are some of the different magnetic foreign objects
that can interact with carbon nanotubes. Among those, transition-metal magnetic adatoms have been reported to produce noticeable changes in the spin-dependent electronic structure of carbon nanotubes\cite{yang03,fazzio2}. Furthermore, defect-induced magnetic moments in carbon-based
materials have been reported as another way to manipulate the magnetic response of these low-dimensional systems\cite{krashenninikov}. 

Motivated by the interaction of carbon nanotubes with foreign magnetic objects, we have recently investigated the problem of magnetic coupling between localized magnetic moments embedded in metallic carbon nanotubes \cite{coupling1,non,c6}. The direct exchange interaction between magnetic moments is known to decay abruptly as a function of their separation since this interaction requires a finite overlap between the wave functions that surround the respective magnetic objects. When the moments are not too close together there is no direct overlap between their wave functions and the only way a magnetic coupling can arise is if it is mediated by the conduction electrons of the metallic host. This so-called indirect exchange coupling (IEC) between localized magnetic moments mediated by the conduction electrons of metallic hosts often plays a central role in determining the overall magnetic order in dilute magnetically-doped metals. 

Put another way, the IEC measures how far apart localized magnetic moments embedded in a metal can feel each other's presence. Metallic nanotubes fall within the category of materials that may display long ranged couplings and have been predicted to display some very interesting features when doped with magnetic objects\cite{coupling1,non,c6}. More specifically, in Reference\cite{coupling1} we have considered how two magnetic adatoms attached to the walls of a nanotube and separated by a distance $D$ are mutually coupled. Surprisingly, we found that the coupling lacks the typical oscillations observed in indirect interaction of magnetic objects embedded in 3-dimensional metallic hosts. We have explained this lack of oscillations as a commensurability effect responsible for hiding the truly oscillatory character of the IEC. Moreover, we have shown that the IEC in nanotubes decays very slowly giving this interaction a very long range. Reference \cite{non} deals with the competition between direct and indirect exchange couplings that occurs when moments are not too far apart and points to the possible existence of non-collinear magnetic order in dimerized impurities. Finally, reference \cite{c6} shows that the long range character of the IEC depends on the exact location of the magnetic moments relatively to the underlying atomic structure of the carbon nanotube. 

All the features mentioned above are associated with the magnetic coupling in a purely static configuration, that is, without accounting for any dynamic aspect of the spin degrees of freedom. Recently, there has been a gradual shift of interest from the static nature of the magnetic interaction towards dynamic aspects of the magnetism of low-dimensional systems, primarily driven by the technological importance of fast magnetization switchings in memory devices. Regarding the magnetic coupling, recent reports indicate the existence of an indirect interaction between precessing magnetic moments separated by a non-magnetic metallic multi-layered material \cite{bauer&cia,bola+mills+bichara}. In particular, precessing magnetic moments that are coupled despite being very far apart suggests that the range of the so-called dynamic coupling always exceeds that of its static counterpart by a considerable amount.  Bearing in mind that the static IEC in nanotubes is already very long ranged \cite{coupling1,non,c6}, it is only natural that we test whether the magnetic coupling can indeed reach further when magnetic moments embedded in nanotubes are allowed to precess. This is the goal of the current manuscript.
For the sake of distinction to earlier descriptions of the dynamic coupling, it is worth mentioning that here we adopt a full quantum mechanical formalism that includes contributions not present in standard semi-classical formalisms, like the one originally carried out in Reference \cite{bauer&cia}. 

The sequence adopted in this manuscript is the following. Next section presents the geometry of the system we will be working with, together with a definition of the corresponding Hamiltonian that describes its electronic structure. The spin susceptibility is a key quantity in this work and is also introduced in the following section. After a physical interpretation is given for the resonant peaks of the diagonal matrix elements of the susceptibility, we shift our interest towards its off-diagonal matrix elements which will be then used to quantify the dynamic coupling throughout this paper. Section  \ref{sec:level2} shows the results of the dynamic coupling between magnetic adatoms for different cases, followed by a brief section with conclusions. 

\section{\label{sec:level1}\protect Model description}

\begin{figure}
\includegraphics[width = 9cm]{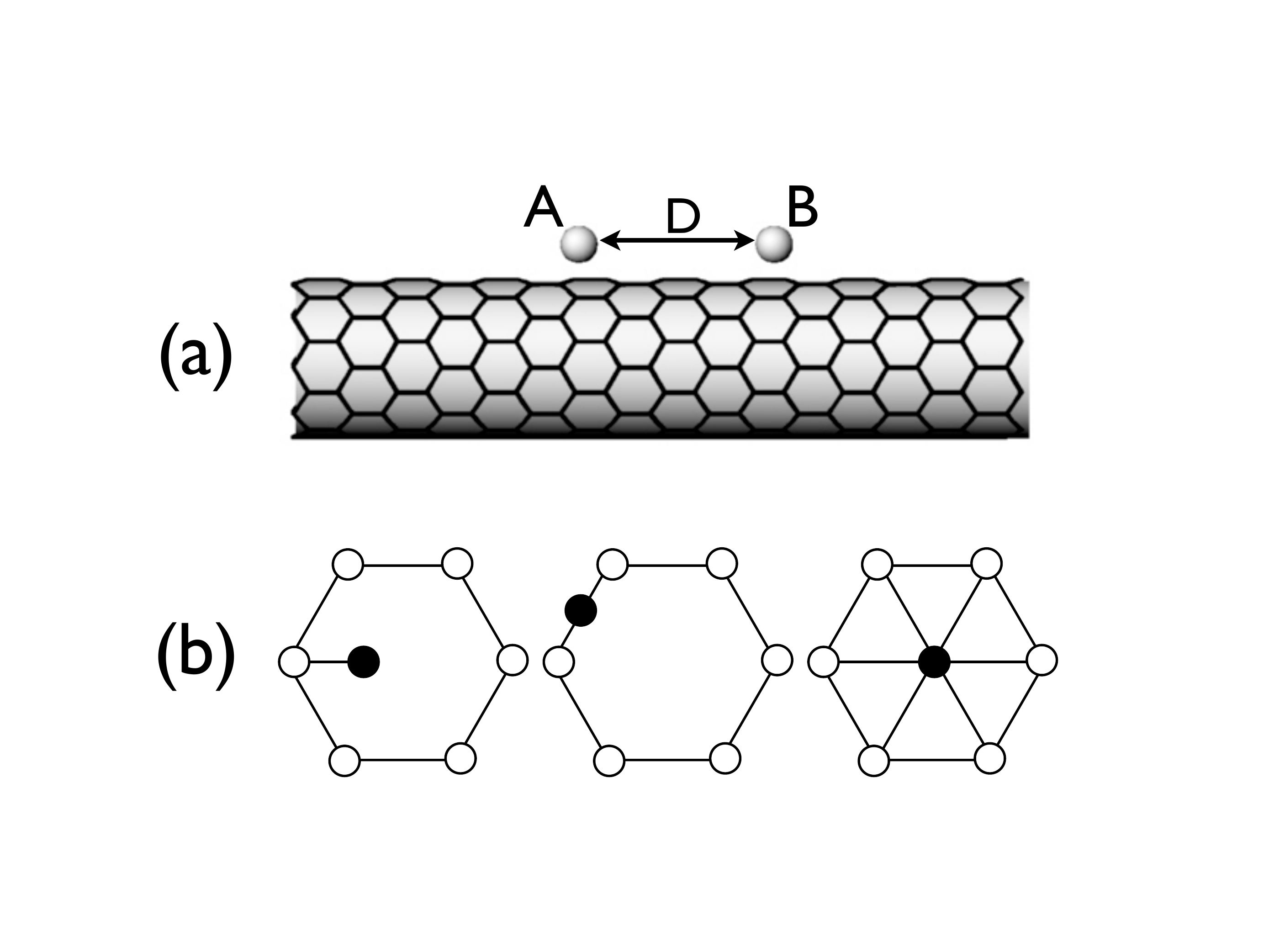}
\caption{(a) Schematic diagram of two magnetic adatoms, labeled A and B, adsorbed to the walls of a carbon nanotube. They are separated by a distance $D$. (b) Possible positions for the magnetic atoms relatively to the underlying hexagonal lattice. On the left, the magnetic atom is adsorbed on top of a single carbon atom, also referred to as the atop position. In the diagram, the adatom has been slightly shifted for clarity. The middle diagram shows the bridge position in which the adatom is above the line joining two nearest-neighbour carbons. Finally, on the right, in the central position, the adatom is above the centre of the underlying hexagon. }
\label{figure_1}
\end{figure}

We consider two magnetic atoms, labelled $A$ and $B$, adsorbed to the walls of an infinitely long metallic carbon nanotube (armchair) and schematically represented in Fig. \ref{figure_1}a. The precise location of these so-called adatoms is one of the three possibilities shown in Fig. \ref{figure_1}b. The first possibility, depicted by the left diagram, corresponds to the adatom being immediately above one carbon atom\cite{comment}. Alternatively, it can lie symmetrically above the bonding line between two neighbouring carbon atoms (middle) or it can be right above the centre of a hexagon composed of six carbon atoms (right). They are labeled as atop, bridge and central positions, respectively. 
It has been recently shown that the range of the static IEC depends on the precise location of the adatom  relatively to the underlying hexagonal lattice\cite{c6}. It is therefore instructive to investigate if and how the dynamic coupling depends on the adatoms location. 

We assume that the magnetic moments of the individual adatoms A and B are initially parallel. We refer to this initial setup as the ferromagnetic configuration, which can be the result of a spontaneous parallel alignment \cite{coupling1}, depending on the adatoms separation, or can be induced by an externally applied uniform magnetic field ${\vec H}$ that drives all the moments to be mutually parallel. Let us now assume that an additional time-dependent magnetic field localized on atom A is applied perpendicularly to the original magnetization direction inducing the moment of adatom A to precess with a frequency $\omega$. Since we are interested in the indirect coupling between the moments A and B, we must investigate the effect that this induced precession on the former will have on the latter. In other words, by calculating the induced motion of the magnetization on adatom B driven by the time-dependent field applied on adatom A, we have a direct way of assessing how strong the indirect coupling between the moments is. 
This is given by the spin susceptibility $\chi(\omega)$, a matrix whose elements reflect how the spin degrees of freedom of the system respond to a magnetic excitation. 

To calculate the spin susceptibility one needs the Hamiltonian operator describing the electronic structure of the system, which is defined by $\hat{H} = \hat{H}_{NT} + \hat{H}_A + \hat{H}_B +\hat{V}_C + \hat{H}_Z$, 
where $\hat{H}_{NT}=\sum_{j,j^\prime,\sigma} \gamma \, {\hat c}_{j,\sigma}^\dag \, {\hat c}_{j^\prime,\sigma}$ is the Hamiltonian of the individual nanotube. The operators ${\hat c}_{m,\sigma}^\dag$ and ${\hat c}_{m,\sigma}$ are the creation and annihilation operators, respectively, for a $\pi$-electron of spin $\sigma = \, \uparrow,\downarrow$ located at site $m$ of the nanotube. The sum over $j$ and $j^\prime$ are over nearest neighbours only and the parameter $\gamma$ stands for the electronic hopping $\gamma = 2.7 {\rm eV}$. ${\hat H}_A$ and ${\hat H}_B$ are Hubbard-like Hamiltonians defined as ${\hat H}_\beta = \sum_\sigma (\epsilon_\beta \, {\hat c}_{\beta,\sigma}^\dag \, {\hat c}_{\beta,\sigma} + {U_\beta \over 2}  \, {\hat n}_{A,\sigma} \, {\hat n}_{A,{\bar \sigma}})$, where $\beta = A,B$ and ${\hat n}_x \equiv {\hat c}_x^\dag {\hat c}_x$ is the number operator. The parameters $\epsilon_\beta$ and $U_{\beta}$ represent the on-site potential and the intra-atomic electronic interaction on the adatom $\beta$, respectively. The operator ${\hat V}_C = \tau \sum_{\ell,\sigma}( {\hat c}_{\ell,\sigma}^\dag {\hat c}_{A,\sigma} +  {\hat c}_{A,\sigma}^\dag {\hat c}_{\ell,\sigma}) +  \tau^* \sum_{\ell^\prime,\sigma}( {\hat c}_{\ell^\prime,\sigma}^\dag {\hat c}_{B,\sigma} +  {\hat c}_{B,\sigma}^\dag {\hat c}_{\ell^\prime,\sigma})$ describes the interaction between the adatoms and the nanotube, where $\tau$ plays the role of an electronic hopping between them. The indices $\ell$ and $\ell^\prime$ label the nanotube sites that are coupled to adatoms $A$ and $B$, respectively. Finally, $\hat{H}_Z$ plays the role of a Zeeman term in the Hamiltonian and is proportional to the applied magnetic field. 

We choose to describe the electronic structure of the system by the single-band tight-binding model, which is justified by the fact that the main features of the IEC are predominantly determined by the extended electrons of the host, in this case the nanotube. As a matter of fact, the electronic
structure of carbon nanotubes is known to be well reproduced by such a model. With transition-metal atoms in mind, the adatoms are also described by a single orbital representing a 5-fold degenerate $d$-band with the appropriate occupation to represent typical magnetic materials. The parameter $\tau$ is selected based on comparisons with {\it ab-initio} evaluations for the static IEC \cite{non} and the parameter $U_\beta$ was assumed to be $1.67 \gamma$ throughout this paper. Despite the simplified model for the electronic structure, our approach is by no means restricted to such a simple case. The results here obtained can be easily extended to a multi-orbital description but bring
no qualitative difference. 

The time-depedent transverse susceptibility is defined as $\chi_{m,n}(t) = -{i \over \hbar} \Theta(t)\langle[{\hat S}_m^+(t),{\hat S}_n^-(0)]\rangle$, where $\Theta(x)$ is the heaviside step function, and ${\hat S}^+$ and ${\hat S}^-$ are the spin raising and lowering operators, respectively. The indices $m$ and $n$ refer to the locations where the field is applied and where the response is measured, respectively. In our case, inducing the moment of adatom $B$ to precess and measuring the corresponding fluctuations of the magnetization on adatom A is described by $\chi_{A,B}(t)$. After Fourier transforming to frequency domain, the susceptibility can be further manipulated and rewritten in matrix form as
\begin{equation}
{\hat \chi}(\omega) = [{\hat 1} + {\hat U} {\hat \chi}^0(\omega)]^{-1} \, {\hat \chi}^0(\omega)\,\,,
\end{equation}
where ${\hat U}$ is a diagonal matrix in which $U_A$ and $U_B$ are the sole elements and ${\hat \chi}^0$ is the Hartree-Fock susceptibility whose matrix elements are
\begin{widetext}
\begin{equation}
\chi_{m,n}^0(\omega) = {i \hbar \over 2 \pi} \int_{-\infty}^{+\infty} d\omega^\prime f(\omega^\prime) \left\{ \left[G_{n,m}^\uparrow(\omega^\prime) -  G_{m,n}^{\uparrow *}(\omega^\prime)  \right]  G_{m,n}^\downarrow(\omega^\prime+\omega) +  \left[G_{m,n}^\downarrow(\omega^\prime) -  G_{n,m}^{\downarrow *}(\omega^\prime)  \right]  G_{m,n}^{\uparrow *}(\omega^\prime+\omega) \right\}\,\,.
\label{susc-hf}
\end{equation}
\end{widetext}
In the equation above $G_{m,n}^\sigma(\omega)$ is the Fourier transform of the time-dependent single-particle Green function ${\cal G}_{m,n}^\sigma(t)$ for an electron with spin $\sigma$ defined as 
\begin{equation}
{\cal G}_{m,n}^\sigma(t) = -{i \over \hbar} \Theta(t) \langle \{{\hat c}_{m,\sigma}(t),{\hat c}_{n,\sigma}^\dag\} \rangle \,\,. 
\end{equation}

In what follows, we show that one way of describing how precessing magnetic moments are aware of each other's presence is by looking at the frequency-dependent spin susceptibility.  Different matrix elements of this quantity reflect distinct features of spin excitations, some of which will be subsequently discussed.

\section{\label{sec:level2}\protect Results}

\begin{figure}
\includegraphics[width = 9cm]{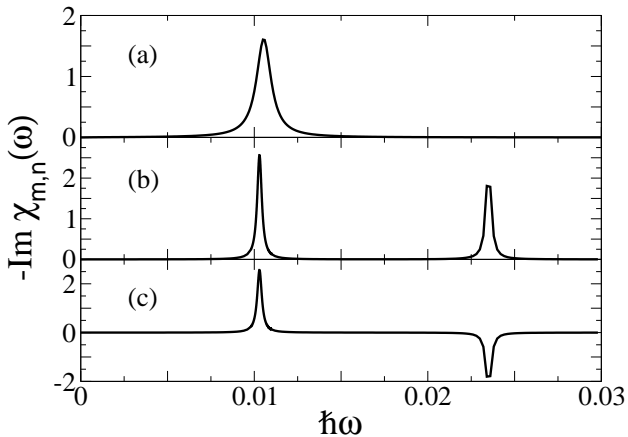}
\caption{Plots of $- {\rm Im} \, \chi_{m,n}(\omega)$ (in arbitrary units) as a function of $\hbar \omega$ (in units of $\gamma$), where $m$ and $n$ label the respective matrix element of the susceptibility. All diagrams are for a (5,5) armchair nanotube and for magnetic atoms described by a d-band occupation of 6 electrons (e.g. Fe). (a) $m=n=A$ corresponds to the diagonal element of the susceptibility calculated for a single magnetic atom ($A$) adsorbed on the central position. (b) Similar matrix element ($m=n=A$) for two magnetic atoms (labeled $A$ and $B$) separated by $D=3a$, where  (c) Off-diagonal matrix element of the susceptibility for the same system as in (b). The difference is in the matrix element, now defined by $m=A$ and $n=B$. }
\label{chi}
\end{figure}

A few characteristic plots of the frequency-dependent spin susceptibility are displayed in Fig.\ref{chi}, whose features require further analysis. The figure shows the imaginary part of matrix elements of the spin susceptibility $\chi(\omega)$. These quantities are directly associated with how strongly the system responds to the time-dependent magnetic field. In other words, large values of $\chi(\omega)$ indicate a greater sensitivity of the system to excitations of frequency $\omega$. In particular, peaks in the $\omega$-dependent susceptibility reflect the existence of resonant frequencies and whose respective widths characterize the lifetime of the spin excitation. In the example depicted on the top panel of Fig. \ref{chi}, the imaginary part of $\chi_{A,A}(\omega)$ corresponding to an armchair nanotube with only a single magnetic adatom (labeled $A$) has a clearly distinctive peak. On the middle panel, ${\rm Im} \, \chi_{A,A}(\omega)$ is calculated for two identical adatoms ($A$ and $B$) a distance $D=3a$ apart along the longitudinal axis of the nanotube, where $a$ is the lattice parameter of the graphene lattice.
Two separate peaks are clearly identifiable and correspond to resonant frequencies associated with the different normal modes which the system can oscillate with, namely the acoustic and the optical modes. In the former the moments of both atoms $A$ and $B$ oscillate in phase but are out of phase in the latter mode. 

Bearing in mind that the peak width is inversely proportional to the characteristic lifetime of a pulsed excitation, it is possible to determine from panels (a) and (b) of Fig. \ref{chi} whether or not the magnetic moments are dynamically coupled. In the case of a single adatom the peak is visibly broader than the corresponding peak (acoustic mode) for the case of two adatoms. This indicates that the precession of atom $A$ is shorter-lived when the adatom is isolated. It is worth highlighting that since there is no spin-orbit interaction in the Hamiltonian, there are no dissipative terms capable of lessening the total angular momentum of the system. This means that the finiteness of the precessional motion lifetime has nothing to do with dissipation but must be related to the way in which the magnetic moments interact with the embedding medium. 

As a matter of fact, we interpret this difference in the excitation lifetimes by suggesting that the precession of the magnetic moment interacts with the surrounding conduction electrons of the nanotube host and induces an angular momentum current, often referred to as spin current\cite{bauer&cia,bola+mills+bichara}, that propagates away from the precessing magnetic moment. In the case of a single magnetic atom, this spin current travels away from the atom with no probability of ever returning to the location where it was originally generated. When two magnetic atoms are present, the spin current emitted by one atom is capable of exciting the precession motion of the other and vice-versa. In this case, the spin current emanating from the atoms is not entirely lost and can be repeatedly scattered back and forth, which  explains why the acoustic mode on Fig.\ref{chi}b is narrower than that of Fig.\ref{chi}a. 

\begin{figure}
\includegraphics[width = 9cm]{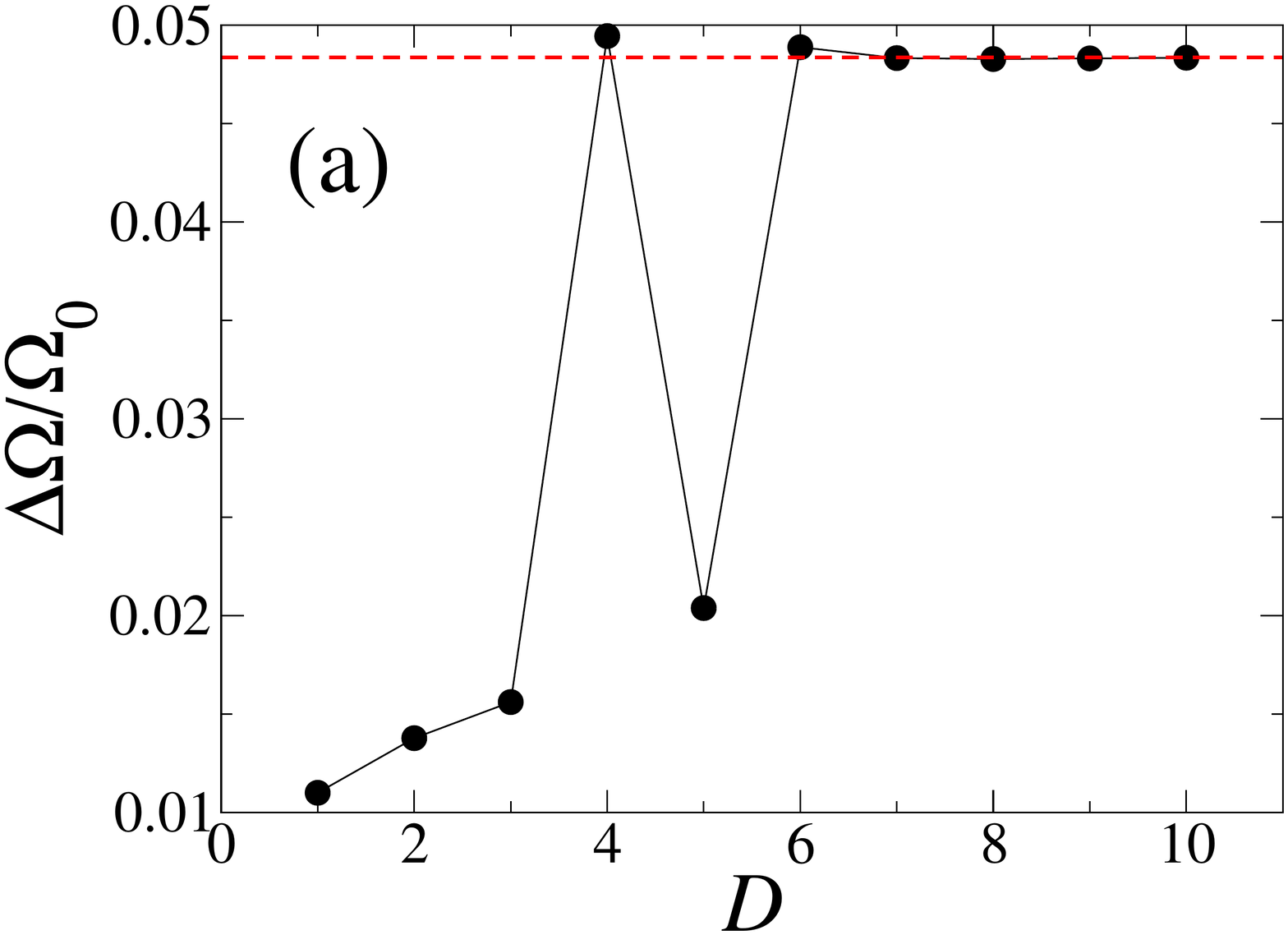}
\includegraphics[width = 9cm]{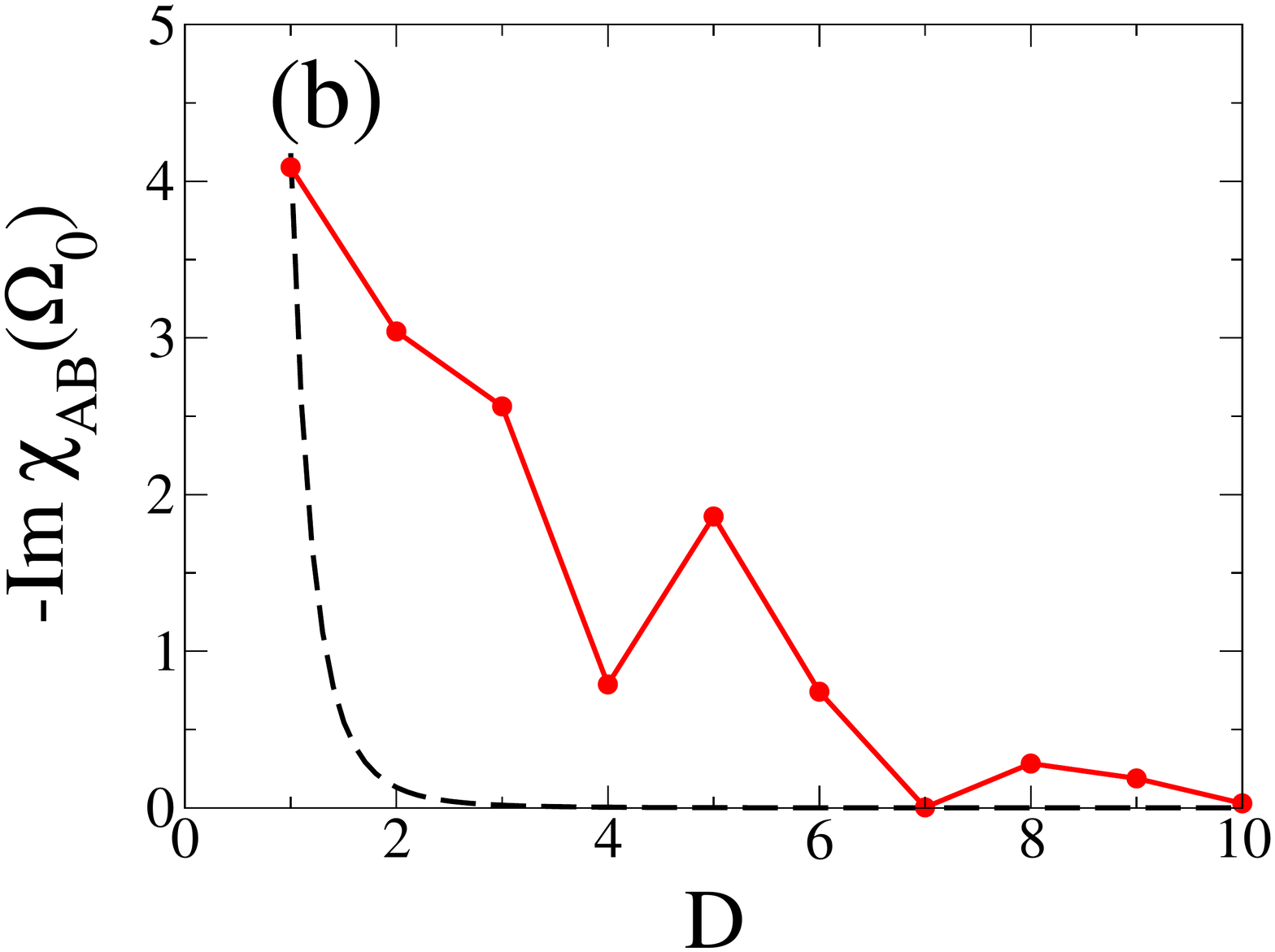}
\caption{Two different ways of inferring about the dynamic coupling. On (a) the width $\Delta \Omega$  of the resonant peak (centred at the acoustic-mode frequency $\Omega_0$) of $- {\rm Im} \chi_{A,A}$ is plotted as a function of the adatom separation $D$ (in units of $a$). When compared with the same quantity obtained for a nanotube with only a single magnetic adatom (represented by the horizontal dashed line), we see that spin excitations of adatoms that are relatively close together live longer than those for which the adatoms are isolated. On (b), similar trends are shown for $-{\rm Im} \chi_{A,B}$ (in arbitrary units), which appears as an alternative quantity displaying the same features. Once again, in both cases the adatoms are located on the central positions and are chosen with the same d-band occupation as those of Fig. \ref{chi}.  }
\label{delta}
\end{figure}

Variations in the width of the acoustic resonant peak are therefore indicative of a dynamic coupling between coexisting magnetic atoms and can in principle be used to probe the existence of this indirect interaction. Fig.\ref{delta}a illustrates how this can be done by plotting $\Delta \Omega/\Omega_0$ for two magnetic adatoms separated by a distance $D$ , where $\Delta \Omega$ is the width of the resonant peak in the acoustic mode and $\Omega_0$ is the precise value of the resonant frequency. The width $\Delta\Omega$ is defined as the energy range around $\Omega_0$ within which the susceptibility exceeds $\chi_{A,A}(\Omega_0)/2$. A horizontal dashed line highlights the corresponding dimensionless quantity for a single impurity. Despite a sudden discontinuity at $D=4a$, reminiscent of the standard oscillatory behaviour found in this type of interaction\cite{bola+mills+bichara}, the general trend is an increase in the width of the acoustic peak towards the limiting value displayed by a single isolated magnetic atom. 

Besides providing a simple physical picture that explains the observed results in the susceptibility calculations, the main reason why the dynamic coupling is analyzed in terms of the widths of the resonating peaks is because this is traditionally how ferromagnetic resonance experiments probe excitations of this nature. Nevertheless, this is unfortunately not the most direct way for evaluating the dynamic coupling between magnetic impurities and an alternative definition is desirable.  

Rather than inferring about the coupling by variations in the width of the resonant peaks, the imaginary part of the off-diagonal matrix element of the spin susceptibility connecting the two magnetic atoms $A$ and $B$ is also a representative quantity for this type of interaction. ${\rm Im}\chi_{A,B}(\omega)$ describes how the moment at atom $A$ responds to an induced excitation of frequency $\omega$ at atom $B$ and should therefore reflect the behaviour seen in Fig.\ref{delta}a in a more direct way. In fact, in spite of not having a clear proportionality relation with the curve in Fig.\ref{delta}a, Fig. \ref{delta}b displays $-{\rm Im} \chi_{A,B}(\Omega_0)$ as a function of $D$ and shows a similar trend to the one reported above in terms of the resonating peak widths, that is, of a coupling that decreases with the atomic separation $D$. Notice that even the discontinuity at $D=4a$ is reproduced. It is thus evident that by associating the dynamic coupling with the imaginary part of $\chi_{A,B}(\Omega_0)$ we obtain a much simpler way of assessing how this interaction depends on $D$. This simplicity is invaluable given the aforementioned motivation of testing whether the range of the dynamic interaction is indeed superior to that of its static counterpart. 

Before discussing how the dynamic coupling that appears in Fig. \ref{delta}b scales with the adatom separation, it is worth mentioning that the static IEC has been reported to decay unusually fast (with approximately $1/D^5$) when magnetic adatoms are located on the central positions of metallic nanotubes\cite{c6}. This fast decaying rate for the static IEC was shown to result from interference effects characteristic of the underlying hexagonal lattice of nanotubes. The dynamic coupling represented by the imaginary part of $\chi_{A,B}$ is depicted by the solid line of Fig. \ref{delta}b and decays considerably more slowly than $1/D^5$. As a matter of fact, starting from $D = a$, the dashed line represents how the dynamic coupling would decay if it was to fall with the same rate as that of the static coupling.  Regarding the solid-line curve of Fig. \ref{delta}b, despite a rapid decay (of above 2 orders of magnitude) between $D=a$ and $D=10a$, this is still far slower than what one would expect from the purely static interaction. 

\begin{figure}
\includegraphics[width = 9cm]{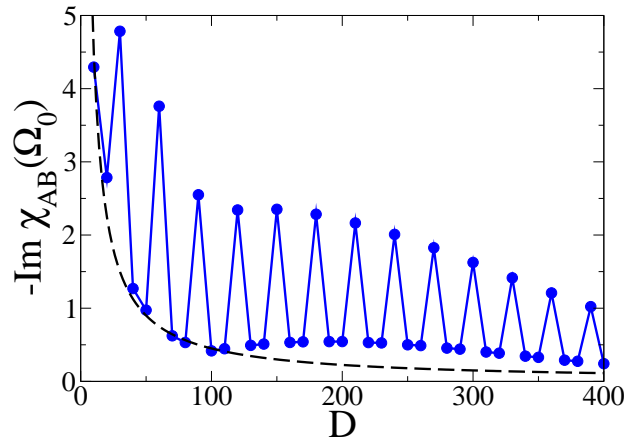}
\caption{$-{\rm Im} \chi_{A,B}(\Omega_0)$ (in arbitrary units) is plotted as a function of the adatom separation $D$, this time for adatoms on the atop position.  $\Omega_0$ corresponds to the resonant frequency of the acoustic mode.  Solid line displays the calculated values for the dynamic coupling. Starting from $D=10a$, the dashed line indicates how the coupling should fall if it was to follow the same rate of decay as the static IEC.}
\label{atop}
\end{figure}
\begin{figure}
\includegraphics[width = 9cm]{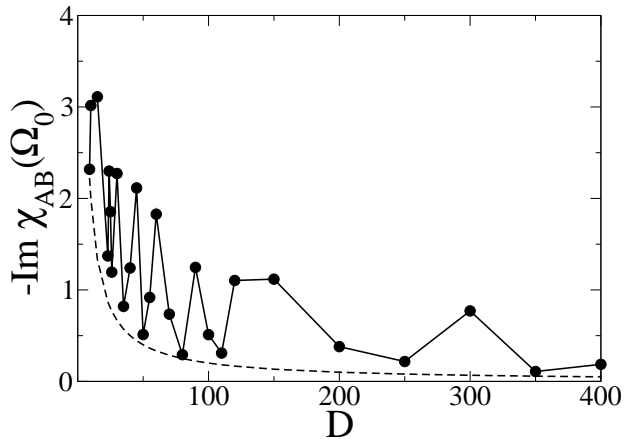}
\caption{$-{\rm Im} \chi_{A,B}(\Omega_0)$ (in arbitrary units) is plotted as a function of the adatom separation $D$, this time for adatoms on the bridge position.  $\Omega_0$ corresponds to the resonant frequency of the acoustic mode.  Solid line displays the calculated values for the dynamic coupling. Starting from $D=9a$, the dashed line indicates how the coupling should fall if it was to follow the same rate of decay as the static IEC. }
\label{bridge}
\end{figure}

For adatoms on both the atop and bridge locations, the static IEC has been shown to decay very slowly with the adatom separation $D$. In fact, the observed $1/D$ decaying rate for these two cases is what characterizes the static IEC as a long-range interaction\cite{c6}. When the dynamic aspects of the indirect coupling are included, the range of this interaction is extended even further, as can be seen in Figs. \ref{atop} and \ref{bridge}. Dashed lines on both graphs indicate the $1/D$ behaviour expected for the static coupling whereas the solid lines show the dynamic coupling through the imaginary part of the off-diagonal matrix element of the spin susceptibility. In both cases, a pronounced oscillatory behaviour is evident but most remarkably, the range of the dynamic coupling is far superior to that of the static IEC. In summary, what was already considered a long range interaction appears to reach even longer distances when the magnetic moments on the walls of a metallic nanotube are put in precessional motion. 

\section{\label{sec:level3}\protect Conclusions}

The direct conclusion from the results presented above is that no matter how long ranged the static IEC is, the dynamic version of this indirect interaction seems to be of superior range. This is in agreement with earlier semi-classically-based predictions that a dynamic coupling may exist even in the absence of its static counterpart\cite{bauer&cia,bola+mills+bichara}.  Here, based on a quantum mechanical approach describing the electronic structure of the system, we are able to confirm this earlier prediction and, moreover, conclude that magnetic moments attached to the walls of a metallic nanotube can feel each other's presence further apart when they are in motion. The fact that setting the moments in motion expands the range of the indirect interaction is quite remarkable, bearing in mind that the static IEC is already very long ranged due to its reduced dimensionality.  

Being this dynamic interaction of such long range, it is likely that this could be exploited as a communication channel between distant parts of spintronic devices. Schemes for synchronizing large sets of oscillators have also been devised based on the dynamic coupling between magnetic 
moments\cite{nature,prb-Fert}. In particular, the ability to excite in-phase oscillations to a magnetic moment that is far apart from the region where the inductive magnetic field is originally produced may enable control of very large arrays of coupled spin-transfer devices opening the road to the construction of macroscopically-sized devices with the above discussed quantum features. 

\begin{acknowledgments}
The authors would like to thank D. F. Kirwan for his assistance with some of the figures in this manuscript. M. S. F. acknowledges the financial support of Science Foundation Ireland. A. T. C acknowledges support received from the brazilian research council, CNPq.    
\end{acknowledgments}


\begin{references}

\bibitem{alphenaar} K. Tsukagoshi, B. W. Alphenaar, and H. Ago, Nature {\bf 401}, 572 (1999).

\bibitem{littlewood} L. E. Hueso, J. M. Pruneda, V. Ferrari, G. Burnel, J. P. Valdes-Herrera, B. D. Simons, P. B. Littlewood, E. Artacho, A. Fert, and N. D. Mathur, Nature {\bf 445}, 410 (2007)

\bibitem{ferreira04} M. S. Ferreira and S. Sanvito, Phys. Rev. B {\bf 69}, 035407 (2004).

\bibitem{cespedes04} O. Cespedes, M. S. Ferreira, S. Sanvito, M. Kociak, and J. M. D. Coey, J. Phys.: Condens. Matter {\bf 16}, 155 (2004).

\bibitem{fazzio1} S. B. Fagan, R. Mota, R. J. Baierle, A. J. R. da Silva, and A. Fazzio, Mater. Charact. {\bf 50}, 183 (2003).

\bibitem{fazzio2} S. B. Fagan, R. Mota, A. J. R. da Silva, and A. Fazzio, Phys. Rev. B {\bf 67}, 205414(2003).

\bibitem{fazzio3} S. B. Fagan, R. Mota, R. J. Baierle, A. J. R. da Silva, and A. Fazzio, Physica B {\bf 340}, 982 (2003).

\bibitem{yang03} C. K. Yang, J. Zhao, and J. P. Lu, Phys. Rev. Lett. {\bf 90}, 257203 (2003)

\bibitem{krashenninikov} Yuchen Ma, A. S. Foster, A. V. Krasheninnikov, and R. M. Nieminen, Phys. Rev. B {\bf 72}, 205416 (2005) 

\bibitem{coupling1} A. T. Costa, D. F. Kirwan, and M. S. Ferreira, Phys. Rev. B {\bf 72}, 085402 (2005)

\bibitem{non} A. T. Costa, C. G. Rocha and M. S. Ferreira, Phys. Rev. B {\bf 76}, 085401 (2007)

\bibitem{c6} D. F. Kirwan, C. G. Rocha, A. T. Costa and M. S. Ferreira, Phys. Rev. B {\bf 77}, 085432 (2008)

\bibitem{bauer&cia} B. Heinrich, Y. Tserkovnyak, G. Woltersdorf, A. Brataas, R. Urban and G. E. W. Bauer, Phys. Rev. Lett. {\bf 90}, 187601 (2003)

\bibitem{bola+mills+bichara}
A. T. Costa, R. B. Muniz and D. L. Mills, Phys. Rev. B {\bf 73}, 054426 (2006)

\bibitem{comment} In the diagram, the adatom has been slightly shifted for clarity.

\bibitem{nature} S. Kaka, M. Pufall, W. Rippard, T. Silva, S. Russek, and 
J. Katine, Nature 437, 7057 (2005) 

\bibitem{prb-Fert} J. Grollier, V. Cros, and A. Fert, Phys. Rev. B 76, 
060409 (2006)

\end{references}
\end{document}